\documentclass[onecolumn,sort&compress,numbers]{els-mrw} 

\usepackage{amsmath,amssymb,amsfonts,amsthm,makeidx,graphicx}
\usepackage{txfonts}
\usepackage{helvet}

\usepackage{cleveref}

\begin{document}


\chapter{Beyond-the-Standard-Model Physics in the Neutrino Sector}\label{bsmnu}

\author[1]{Kevin J. Kelly}%

\address[1]{\orgname{Texas A\&M University}, \orgdiv{Department of Physics \& Astronomy, Mitchell Institute for Fundamental Physics and Astronomy}, \orgaddress{4242 TAMU, College Station, TX 77845}}

\articletag{Chapter Article tagline: BSM Physics with Neutrinos.}

\maketitle

\begin{abstract}[Abstract]
	Neutrino oscillations are a phenomenon that has been observed for over two decades and leads to the conclusion that neutrinos have mass.
	The Standard Model predicts massless neutrinos, and so neutrinos require physics beyond the Standard Model.
	Other signatures of BSM physics are detectable in modern neutrino facilities -- this chapter explores those possibilities.
	These can range from new effects modifying neutrino oscillations (beyond the expectations when neutrinos have mass), to searches for new particles in neutrino facilities.
	Next-generation experiments are particularly powerful for these searches due to high-intensity neutrino beams and novel detection technologies.
	We give an introduction to these search strategies, giving a non-comprehensive overview of the field as it stands presently.
\end{abstract}

\begin{keywords}
 	neutrinos\sep BSM\sep sterile neutrinos \sep non-standard neutrino interactions \sep dark sectors
\end{keywords}

\section*{Objectives}
\begin{itemize}
	\item Understand how neutrino oscillations are measured and how certain classes of new physics can impact this phenomenon.
	\item Explore how neutrino near detectors are well-suited to search for new physics from neutrinos and otherwise.
	\item Envision neutrino detectors as machines capable of searching for particles associated with the dark matter in the universe.
\end{itemize}

\section{Introduction}\label{intro}

Neutrinos, as a fundamental particle, have gained significant interest over the last two decades with the discovery of the phenomenon of neutrino oscillations. This arises from the fact that neutrinos have mass and that there is (sizable) mixing between the states by which the neutrinos interact and by which they propagate -- this fact itself requires physics beyond the Standard Model. Furthermore, modern facilities are operating and under development to precisely study neutrino oscillations and neutrino interactions. Thanks to these high-precision facilities, we are enabled in searching for copious types of beyond-the-Standard-Model physics.

This chapter explains modern research focused on these strategies, dividing the discussion roughly into two parts. First, we explore the phenomenon of neutrino oscillations and how additional new physics can modify our measurements from the expected ``standard'' picture. New physics of this type includes the potential existence of light sterile neutrinos and novel neutrino interactions with matter, often referred to as ``non-standard'' neutrino interactions. We also discuss how high-statistics facilities designed to study neutrino interactions may search for rare processes -- in turn, these rare processes are sensitive to corrections from new physics, for instance the Standard Model neutrino-trident interaction being impacted by the presence of a new, light, neutrinophilic gauge boson. As a second class of models, we explore how neutrino facilities can be sensitive to new physics not at all connected to neutrinos. The main example of this that we discuss is the possible existence of dark sectors, a set of one or more new-physics particles that can be detected in these high-intensity facilities. Models of this nature include light dark matter, coupled to the Standard Model through a new mediator such as a dark photon, or heavier metastable states such as heavy neutral leptons.

The aim of this chapter is to introduce the reader to these very exciting search strategies and some of the open questions currently being addressed. References throughout should allow the interested reader to explore deeper and discover alternative search strategies and to investigate additional open questions not yet explored in the literature.

\section{Deviations from standard neutrino oscillations}\label{oscillationBSM}
When the three SM neutrinos have distinct masses $m_1$, $m_2$, and $m_3$, there is a mismatch between this mass basis and the basis in which neutrinos interact via the SM weak interactions -- the flavor basis.
This mismatch is expressed using a unitary, leptonic mixing matrix $U$, defined by
\begin{equation}
\nu_\alpha = U_{\alpha i} \nu_i, \quad \lvert \nu_\alpha \rangle = U_{\alpha i}^* \lvert \nu_i\rangle,
\end{equation}
where the first portion refers to the neutrino fields and the latter to the neutrino states, hence the difference in $U \to U^*$. 
The matrix $U$, if unitary and $3\times3$, requires three mixing angles and one Charge-Parity (CP)-violating phase to which neutrino oscillations are sensitive. In the standard ``PDG'' parameterizatoin, $U$ takes the form
\begin{align}
U &= \left(\begin{array}{c c c} 1 & 0 & 0 \\ 0 & c_{23} & s_{23} \\ 0 & -s_{23} & c_{23} \end{array}\right) \left(\begin{array}{c c c} c_{13} & 0 & s_{13} e^{-i\delta_{\rm CP}} \\ 0 & 1 & 0 \\ -s_{13} e^{i\delta_{\rm CP}} & 0 & c_{13} \end{array}\right) \left(\begin{array}{c c c} c_{12} & s_{12} & 0 \\ -s_{12} & c_{12} & 0 \\ 0 & 0 & 1\end{array}\right), \\
&= \left(\begin{array}{c c c} c_{12} c_{13} & s_{12} c_{13} & s_{13} e^{-i\delta_{\rm CP}} \\ -s_{12} c_{23} - c_{12} s_{23} s_{13} e^{i\delta_{\rm CP}} & c_{12} c_{23} - s_{12} s_{23} s_{13} e^{i\delta_{\rm CP}} & s_{23} c_{13} \\ s_{12} s_{23} - c_{12} c_{23} s_{13} e^{i\delta_{\rm CP}} & -c_{12} s_{23} - s_{12} c_{23} s_{13} e^{i\delta_{\rm CP}} & c_{23} c_{13} \end{array}\right),
\end{align}
where $s_{ij} = \sin\theta_{ij}$ and $c_{ij} = \cos\theta_{ij}$. The three mixing angles $\theta_{12}$, $\theta_{13}$, and $\theta_{23}$ have been measured to various levels of precision, and experiments have begun to constrain the allowed range of the complex phase $\delta_{\rm CP}$.

The time-evolution of the neutrinos propagating through SM matter is expressed by
\begin{align}\label{eq:HamiltonianStandard}
H_{ij} = \frac{1}{2E_\nu} \mathrm{diag} \left\lbrace 0, \Delta m_{21}^2, \Delta m_{31}^2\right\rbrace + V_{ij},
\end{align}
where $\Delta m_{ji}^2 \equiv m_j^2 - m_i^2$ expresses the \textit{mass-squared-differences} of the neutrino mass eigenstates and $V_{ij}$ includes the potential sourced by neutrino/SM interactions.
We provide the Hamiltonian in the mass basis, i.e.~$H_{ij}$, as opposed to the flavor-basis Hamiltonian $H_{\alpha\beta}$ for simplicity.
The potential term $V_{ij}$ can be expressed as
\begin{align}
V_{ij} &= U_{i\alpha}^\dagger V_{\alpha\beta} U_{\beta j}, \\
V_{\alpha\beta} &= \mathrm{diag}\left\lbrace A, 0, 0\right\rbrace, \label{eq:Vab}
\end{align}
where $A = \sqrt{2}G_F n_e$, $G_F$ is the Fermi constant and $n_e$ is the electron number density along the path of propagation.
The fact that the only nonzero term in the flavor-basis $V_{\alpha\beta}$ is in the $(ee)$ component is because all neutrino flavors interact with the same cross section with electrons, protons, and neutrons via $Z$-boson exchange, but electron neutrinos have an additional contribution when scattering with $e^-$ due to charged $W^\pm$ exchange.

With these ingredients, we may calculate the oscillation probability that a neutrino produced as a flavor $\alpha$ travels a distance $L$ and is detected as a state with flavor $\beta$,
\begin{align}
P\left(\nu_\alpha \to \nu_\beta\right) &= \left\lvert \langle \nu_\beta \rvert U e^{-iH_{ij} L} U^\dagger \lvert \nu_\alpha \rangle \right\rvert^2.
\end{align}
Effectively, we rotate the initial-state $\nu_\alpha$ into the mass basis, apply a time-evolution operator, and rotate back to the flavor-basis, projecting on to the final-state $\nu_\beta$.
The oscillation probabilities for channels of interest, especially among $\alpha,\beta$ = $e$ and/or $\mu$, can be calculated exactly for three-neutrino oscillations propagating in vacuum.
Compact expressions for oscillations through constant matter (e.g. long-baseline oscillations through the earth) can be constructed using appropriate perturbative approaches, see, e.g., Refs.~\cite{Denton:2016wmg,Barenboim:2019pfp}.

Working within the three-neutrino framework, measurements of oscillation probabilities allow us to better understand the mixing matrix $U$ (and the mixing angles/CP-violating phase within) as well as the mass-squared differences $\Delta m_{ji}^2$. Canonically, six parameters control neutrino oscillations. The bulk of these are measured at the precision level, see, e.g., Refs.~\cite{deSalas:2020pgw,Capozzi:2021fjo,Esteban:2024eli} for summaries of our current understanding of three-neutrino oscillation.

Now we are prepared to study how beyond-the-standard-model physics may modify this picture. Many scenarios and models exist which impact neutrino oscillations, and here we will focus on two particular classes:
\begin{enumerate}
\item One or more new states that mix with neutrinos impact their mixing and/or oscillation.
\item New interactions between neutrinos and SM particles modify their propagation.
\end{enumerate}
We will explore these two classes in detail in the following subsections.

\subsection{New neutrino state(s)}\label{subsec:NewStates}
The Hamiltonian in~\cref{eq:HamiltonianStandard} assumes that three light neutrinos exist, and that the mixing matrix $U_{\alpha i}$ characterizes rotations between the three mass eigenstates and the three flavor eigenstates associated with charged leptons $e$, $\mu$, and $\tau$. While copious studies restrict the existence of new charged-fermion states, at least below the electroweak scale, there is the possibility that new, light, SM gauge-singlet fermions exist. If they do, they share the same quantum numbers, after electroweak symmetry breaking, as the SM neutrinos and therefore, these states can mix.

If there are $3 + N$ ($N \geq 1$) neutrino states, then the mixing matrix $U_{\alpha i}$ is extended from being $3\times 3$ to $(3+N)\times (3+N)$, while still being unitary.\footnote{Typically the matrix is taken to be square to apply unitarity, however the elements along the ``sterile'' rows (those rows beyond $e$, $\mu$, $\tau$) are in principle unphysical.} In the case where the new sterile states are very heavy, the only imprint is that the $3\times 3$ sub-matrix of $U_{\alpha i}$, to which conventional oscillation experiments would be sensitive, is itself no longer unitary. Searches for deviations from unitarity provide a powerful, yet general, method of testing whether global data are consistent with the standard three-neutrino framework~\cite{Antusch:2006vwa,Antusch:2014woa,Li:2015oal,Blennow:2016jkn,Parke:2015goa,Ellis:2020hus}.

Let us turn to the case where, for concreteness, $N = 1$ (one additional light sterile neutrino) and its associated mass $m_4$ is such that it can travel coherently with the other light neutrinos $\nu_{1,2,3}$ (for discussion of coherent wave-packet propagation of neutrino states including sterile neutrinos, see Ref.~\cite{Jones:2014sfa}). As mentioned, the mixing matrix is now extended -- one additional sterile neutrino requires the addition of three mixing angles $\theta_{14}$, $\theta_{24}$, and $\theta_{34}$, as well as two additional CP-violating phases $\delta_{24}$ and $\delta_{34}$. A bevy of experiments perform dedicated searches for coherent oscillations between the SM neutrinos and this sterile neutrinos, searching for evidence of nonzero mixing angles $\theta_{i4}$ as a function of the new mass-squared splitting $\Delta m_{41}^2$.

Due to their gauge-singlet nature, the ``sterile'' flavor (i.e.,\ the fourth linear combination of the four mass eigenstates other than $e$, $\mu$, and $\tau$) does not interact with SM matter as it propagates. This requires a correction to the matter potential in~\cref{eq:Vab}, since we assumed uniform neutral-current interactions of the flavor-basis neutrinos with electrons, protons, and neutrons along the neutrinos' path of propagation. In this case, the flavor-basis matter potential is modified,
\begin{align}
V_{\alpha\beta} \to \mathrm{diag}\left\lbrace A, 0, 0, A^\prime\right\rbrace,
\end{align}
where $A^\prime = A/2$ for oscillations in neutral, isoscalar matter. This difference can manifest large differences to the neutrino oscillation probabilities at long distances, as explored in Ref.~\cite{deGouvea:2022kma}.

Generically, neutrino oscillation experiments are capable of searching for the deviations imprinted by these new, light sterile neutrinos on the oscillation channels that they measure. For example, the NOvA experiment, through its charged-current ($\nu_\mu \to \nu_\mu$ and $\nu_\mu \to \nu_e$) channels, as well as its neutral-current one, is capable of precisely constraining $\sin^2\theta_{24}$ and $\sin^2\theta_{34}$ in the range $10^{-3}$ eV$^2 < \Delta m_{41}^2 < 10^2$ eV$^2$~\cite{NOvA:2017geg,NOvA:2021smv,NOvA:2024imi}. These results arise from the comparison of expected event rates to actual data in both NOvA's near and far detectors. Similarly, the T2K experiment uses its near and far detectors and all possible oscillation channels to constrain the existence of sterile neutrinos across a wide range of $\Delta m_{41}^2$ in Ref.~\cite{T2K:2019efw}.

\subsection{New neutrino interactions}\label{subsec:NSI}
When constructing the oscillation formalism above, we not only assumed that only three neutrinos exist, we also assumed that their interactions were exactly as prescribed by the SM, i.e.\ that their interactions with SM charged fermions arise exclusively from the exchange of $W$- and $Z$-bosons. Through these interactions with electrons, protons, and neutrons, this generates the matter potential for neutrinos along their path of propagation as expressed in~\cref{eq:Vab}. Given that neutrinos are much less well-understood than their charged-fermion peers in the SM, it remains possible that new interactions among the neutrinos, as well as between the neutrinos and SM particles, exist. This is typically formulated in the context of ``non-standard neutrino interactions'' (NSI) that can modify any/all of the production, propagation, and detection of the neutrinos.

First, let us briefly discuss how NSI can impact the production and/or detection of neutrinos. If new four-fermion operators involving neutrinos and charged leptons exist, they can take, for example, the form
\begin{align}
\mathcal{L} \supset \frac{1}{\Lambda^2}\left(\bar{\nu}_\alpha P_L e\right)\left( \bar{f} \left(c_L P_L + c_R P_R\right) f^\prime\right) + \mathrm{h.c.},
\end{align}
where $c_L$ and $c_R$ are dimensionless coefficients, $\Lambda$ is the scale of new-physics associated with this dimension-six operator, and $f$ and $f\prime$ represent two SM fermions. Such an operator would lead to new methods of neutrino charged-current scattering, potentially including apparent flavor violation if $\alpha \neq e$, allowing muon- or tau-neutrinos to produced charged-electrons. Similar charged-current new-physics operators can impact SM charged meson decays and provide new, nonstandard production of neutrinos that can be searched for in neutrino-beam facilities.

For the remainder of this subsection, we will explore the possibility that NSI exist only in the form of new neutral-current interactions, which (predominantly) impact the propagation of neutrinos. These take the form
\begin{align}\label{eq:NSI}
\mathcal{L} \supset -2\sqrt{2}G_F \left( \bar{\nu}_\alpha \gamma_\rho \nu_\beta\right) \left( \epsilon_{\alpha\beta}^{f \tilde{f},L} \bar{f}_L \gamma^\rho \tilde{f}_L + \epsilon_{\alpha\beta}^{f \tilde{f},R} \bar{f}_R \gamma^\rho \tilde{f}_R\right) + \mathrm{h.c.},
\end{align}
characterizing new interactions between neutrinos of flavor $\alpha$ and $\beta$ with SM fermions $f$ and $\tilde{f}$. This prescription expresses the new interactions relative to the strength of the weak interaction, hence the appearance of $G_F$ so that $\epsilon_{...}^{...}$ are dimensionless constants.

In addition to their impact on neutrino propagations, these neutral-current NSI can be searched for in processes involving neutrino neutral-current scattering, for instance the coherent-scattering CEvNS process detected by the COHERENT collaboration~\cite{COHERENT:2017ipa,COHERENT:2020iec}. Data from COHERENT have been analyzed in the context of neutral-current NSI in, for example, Refs.~\cite{COHERENT:2017ipa,Denton:2018xmq,Giunti:2019xpr,Denton:2022nol}. More recently, the first evidence of solar neutrinos interacting with dark matter direct-detection experiments through the CEvNS process was reported by the PandaX-4T~\cite{PandaX:2024muv} and XENONnT~\cite{XENON:2024ijk} collaborations -- this has been analyzed in the context of neutral-current NSI in Ref.~\cite{AristizabalSierra:2024nwf}.

The existence of nonzero $\epsilon_{...}^{...}$ in~\cref{eq:NSI} lead to novel interactions of neutrinos with SM matter along the path of propagation, modifying the matter potential discussed above. Summing over all potential interactions (first, assuming $f = \tilde{f}$ for coherent interactions, as well as $\epsilon_{\alpha\beta}^f \equiv \epsilon_{\alpha\beta}^{f,L} + \epsilon_{\alpha\beta}^{f,R}$ for vector-like interactions), we can determine the effective parameters impacting neutrino oscillations,
\begin{align}
\epsilon_{\alpha\beta} = \sum_{f=u,d,e} \epsilon_{\alpha\beta}^{f} \frac{n_f}{n_e},
\end{align}
where $n_f$ is the number density of fermion $f$ -- we assume that for neutral, isoscalar matter, $n_u/n_e = n_d/n_e = 3$. This allows us to express the updated neutrino matter potential (in the flavor basis) as
\begin{align}
V_{\alpha\beta} \longrightarrow A\left(\begin{array}{c c c} 1 + \epsilon_{ee} & \epsilon_{e\mu} & \epsilon_{e\tau} \\ \epsilon_{e\mu}^{*} & \epsilon_{\mu\mu} & \epsilon_{\mu\tau} \\ \epsilon_{e\tau}^{*} & \epsilon_{\mu\tau}^{*} & \epsilon_{\tau\tau}\end{array}\right).
\end{align}
Here, $A$ is still equal to $\sqrt{2}G_F n_e$ as in the standard case, and the expression reduces to the SM in the limit where all $\epsilon_{\alpha\beta} \to 0$. This parameterization includes the addition of six real parameters (the real diagonal parameters and the magnitudes of the three off-diagonal ones) as well as three complex phases associated with the three off-axis parameters.

The impact of non-standard interactions on long-baseline oscillations has been studied in a vast array of scenarios in the literature -- we refer the reader to Ref.~\cite{Proceedings:2019qno} for a review. Examples of upcoming experimental sensitivity to non-standard neutrino interactions via their oscillation can be found in Refs.~\cite{deGouvea:2015ndi,Coloma:2015kiu,Kelly:2017kch}. Recently, non-standard interactions with sizeable CP violation (via the aforementioned complex phases for off-diagonal entries) have been invoked to explain mild tensions between NOvA~\cite{NOvA:2021nfi,NOvA:2023iam} and T2K~\cite{T2K:2023smv} oscillation analyses in Refs.~\cite{Denton:2020uda,Chatterjee:2020kkm} -- such explanations should be tested thoroughly in the coming generation of oscillation experiments~\cite{Denton:2022pxt}.

\paragraph{Scalar Non-standard Neutrino Interactions in Propagation}
Until recently, the bulk of analyses regarding neutral current non-standard neutrino interactions studied the vector-like interactions, akin to the SM $Z$-boson exchange processes. As identified in Ref.~\cite{Ge:2018uhz}, interactions between neutrinos and SM fermions via a new scalar particle yield a qualitatively different impact on neutrino oscillations. Such a scalar interaction ends up modifying the Dirac equation of neutrino evolution, proportional to the strength of the new interaction and the number density of the ``target'' particles (electrons, up quarks, and down quarks in the case of oscillations through standard matter). The scalar-mediated interaction leads to a number-density dependent induced neutrino mass that interferes with the neutrinos' vacuum masses as they propagate. Such new interactions will naturally be explored as neutrino oscillation data explore the precision era in the coming decades.

\section{New physics with neutrinos near sources}\label{nearBSM}
In~\cref{oscillationBSM}, we focused on studying neutrino oscillations and exploring deviations from the standard three-neutrino framework as a mechanism of searching for new physics. Here, we focus on situations in which neutrino sources and detectors are relatively close, such that no SM oscillations are expected on these distance and energy scales. The proximity of the sources to the detectors can allow for much larger event rates than in the oscillation-focused experiments, and such large-statistics searches can afford precision studies that are sensitive to novel beyond-the-SM mechanisms.

One such mechanism was discussed in~\cref{subsec:NSI} via the presence of charged-current non-standard neutrino interactions. Such new physics would lead to modifications from the expected production and detection of neutrinos, and high-statistics experiments provide the most stringent tests of such new physics. These effects have been studied in an effective-field-theory approach in various intense-neutrino environments, ranging from reactors to neutrinos produced in hadron colliders, in Refs.~\cite{Falkowski:2018dmy,Falkowski:2019xoe,Falkowski:2021bkq}.

In the remainder of this section, we explore a handful of concrete scenarios that can be tested well in such environments. In~\cref{subsec:WeakAngle}, we discuss how the standard model weak mixing angle, $\theta_w$ can be measured using neutrino interactions.~\cref{subsec:RareProcesses} focuses on standard model processes which are rare (albeit detectable), and mechanisms by which new physics may interfere constructively or destructively with these processes. Finally,~\cref{subsec:Reactors} focuses on measurements of neutrinos produced in nuclear reactors and ways in which these low-energy neutrinos are particularly useful in searches for new physics.

\subsection{Precision weak-interaction measurements}\label{subsec:WeakAngle}
In the standard model, the weak interactions' strength (and particles' couplings to the $W$- and $Z$-bosons) are governed by the weak mixing angle $\theta_W$. Famously, the weak mixing angle is not a constant, but it varies depending on the momentum-transfer scale $q$ at which it is measured. Numerous experiments have measured this running across scales between $10^{-3}\sim 100$~GeV, using a wide variety of experimental conditions to do so.

Neutrino scattering is particularly sensitive to $\theta_W$, particularly in using neutrino-electron elastic scattering. Neutrino-electron scattering obeys a relatively small cross section (about a factor of $1000$ smaller than neutrino-nucleus scattering), and so high-intensity environments are necessary in order to precisely measure this process. The upcoming DUNE near detector complex offers such intensities, where thousands of neutrino-electron scattering events are expected to be measured. Refs.~\cite{deGouvea:2019wav} and~\cite{Alves:2024twb} have explored the capabilities of DUNE and SBND respectively in such a measurement. In both cases, these neutrino facilities are able to measure $\theta_W$ in novel regimes of momentum transfer $q$. Such measurements will add to the complementarity of different probes, and, in the event of an inconsistent extraction of $\theta_W$, may point to additional new physics at play in the neutrino sector.

\subsection{Rare processes in the SM}\label{subsec:RareProcesses}
With intense neutrino sources in close proximity to neutrino detectors, it is feasible to search for processes that exist within the SM but with relatively small cross sections. This includes, for instance, the CEvNS process detected by COHERENT~\cite{COHERENT:2017ipa,COHERENT:2020iec} and the neutrino-trident scattering process, in which a neutrino scatters off a nucleus and emits a pair of oppositely charged leptons $\ell_\alpha^+ \ell_\beta^-$. The neutrino-trident cross section is small ($\sigma \approx 10^{-42}$ cm$^2$ for GeV-scale neutrino energies, in contrast with $10^{-38}$ cm$^2$ for charged-current quasielastic scattering) and therefore has only been detected in a handful of experimental facilities~\cite{CHARM-II:1990dvf,CCFR:1991lpl,NuTeV:1998khj}.

Both of these detection processes (and others) benefit in searches for new physics from the smallness of the SM cross sections. They also benefit from the possibility that new physics can interfere coherently with the SM at the matrix-element level, allowing for searches to test much of the new-physics parameter space. As a concrete example, let us briefly explore neutrino-trident scattering with a new mediator $Z^\prime$.
\begin{figure}[!htbp]
	\centering
	\includegraphics[width=0.8\textwidth]{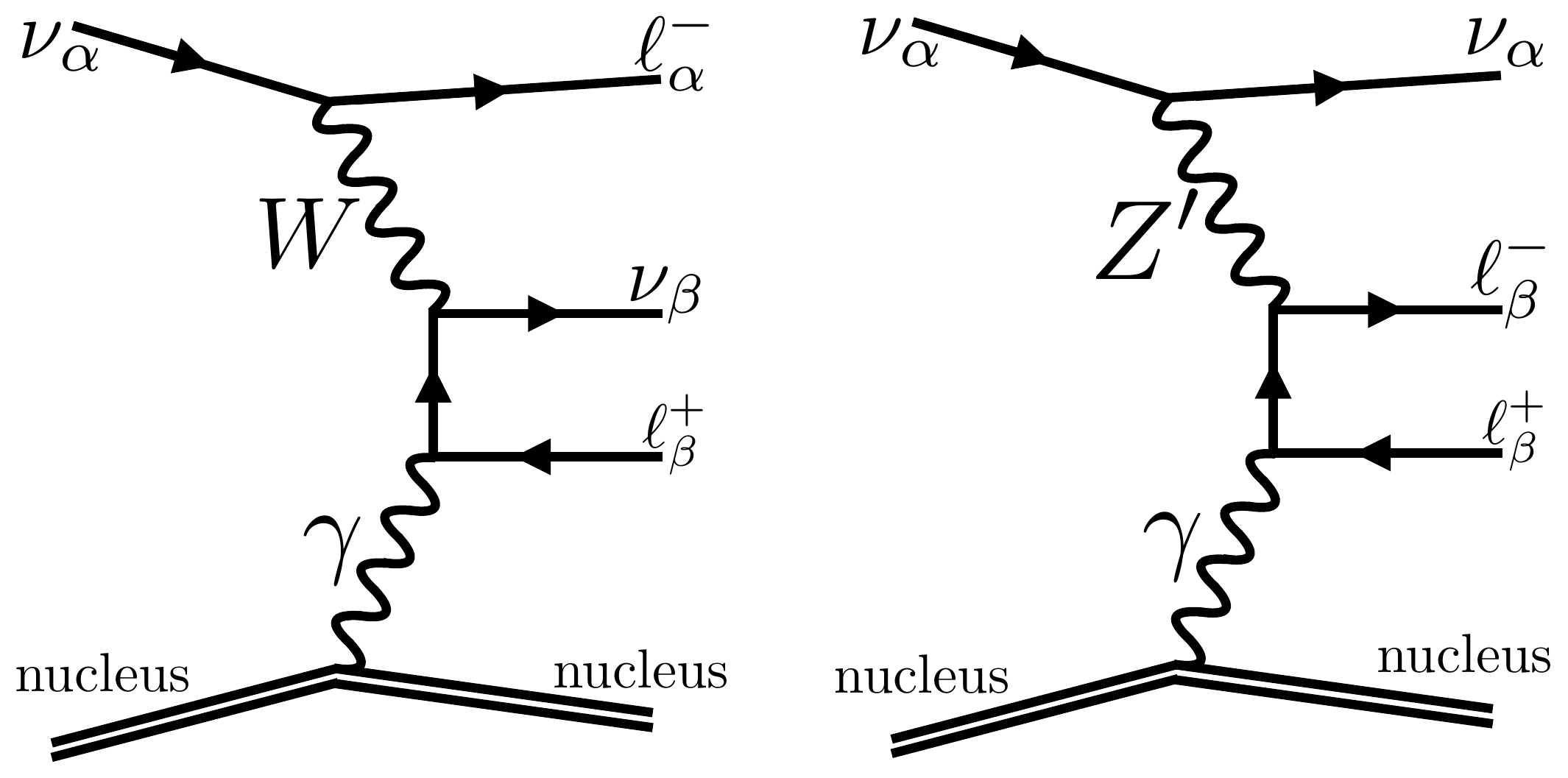}
	\caption{Two characteristic diagrams contributing to neutrino trident scattering, (left) in the Standard Model via $W$-boson exchange, and (right) with the inclusion of a new $Z^\prime$ gauge boson that couples to neutrinos and charged leptons.}
	\label{fig:NeutrinoTrident}
\end{figure}

In~\cref{fig:NeutrinoTrident}, we display two characteristic diagrams that contribute to standard-model (left) and BSM (right) neutrino-trident scattering. In the SM, exchange of $W$- and/or $Z$- (depending on the flavor of the outgoing charged leptons and neutrino) bosons leads to this process. If a new $Z^\prime$ exists, then a diagram such as the right one of~\cref{fig:NeutrinoTrident} exists, where the (presumed-to-be-small) coupling enters at the neutrino-neutrino-$Z^\prime$ vertex. Depending on the outgoing particles' flavors, these two diagrams interfere at the matrix-element level, leading to an overall cross section that scales linearly with the new, small coupling rather than the coupling squared. Additionally, if $Z^\prime$ is light relative to the SM $W/Z$-bosons, the contribution arising from its exchange is enhanced. This constructive interference allows for future experiments, such as the DUNE near detector complex~\cite{Ballett:2018uuc}, to search for light $Z^\prime$ at ${\sim}10-100$~MeV masses for couplings on the order of $10^{-3}$~\cite{Altmannshofer:2019zhy,Ballett:2019xoj} -- a region of particular interest motivated by the long-standing $(g-2)$ anomaly of the muon~\cite{Muong-2:2006rrc,Muong-2:2021ojo,Muong-2:2023cdq,Baek:2001kca}.

\subsection{Precision tests at nuclear reactors}\label{subsec:Reactors}
Nuclear reactors are also provide high-intensity neutrino sources, including serving as the first source of neutrinos ever detected~\cite{Cowan:1956rrn}. In modern days, reactor-antineutrino measurements are an attractive way of searching for new physics due to the large event rates afforded by the inverse-beta-decay scattering process, $\bar\nu_e p \to e^+ n$. Many of the new-physics scenarios discussed above are best tested in studies of reactor antineutrinos.

For instance, if a light sterile neutrino exists and mixes with electron-flavor via nonzero $|U_{e4}|^2$, then reactor antineutrino spectra should have measurable energy-dependent oscillation probabilities $P(\bar\nu_e \to \bar\nu_e)$ at energy and distance scales dictated by the new mass-squared splitting $\Delta m_{41}^2$. For the mass-squared splittings indicated by a sterile neutrino interpretation of the LSND and MiniBooNE excesses, reactor neutrinos measured at $\mathcal{O}(10~\mathrm{m})$ distances are expected to contribute significantly to potential discovery. The current interpretation is that there is insufficient evidence for reactor antineutrino disappearance for a consistent sterile-neutrino interpretation of all of this data -- see, e.g., Refs.~\cite{Giunti:2021kab,Berryman:2021yan,Hardin:2022muu,Dentler:2018sju,Giunti:2019aiy} for thorough discussions.

\section{Dark sector searches with neutrino facilities}\label{neutrinoDarkSectors}
Complementary to the new-physics searches discussed in~\cref{oscillationBSM,nearBSM}, a good deal of recent research in the literature focuses on new physics \textit{not} related to neutrinos that can be searched for in neutrino facilities. Much of this research studies the possibility of producing new particles alongside the bulk of the neutrino beam, and detecting these new particles using one many different strategies. These approaches mirror the type of new-physics and dark-sector searches present in fixed-target, beam-dump facilities over the last several decades. With modern-day neutrino detectors being employed in these beam-dump-esque facilities, we may gain on previous searches due to the excelled particle identification and energy reconstruction capabilities of these detectors.

\begin{figure}[!htbp]
	\centering
	\includegraphics[width=\textwidth]{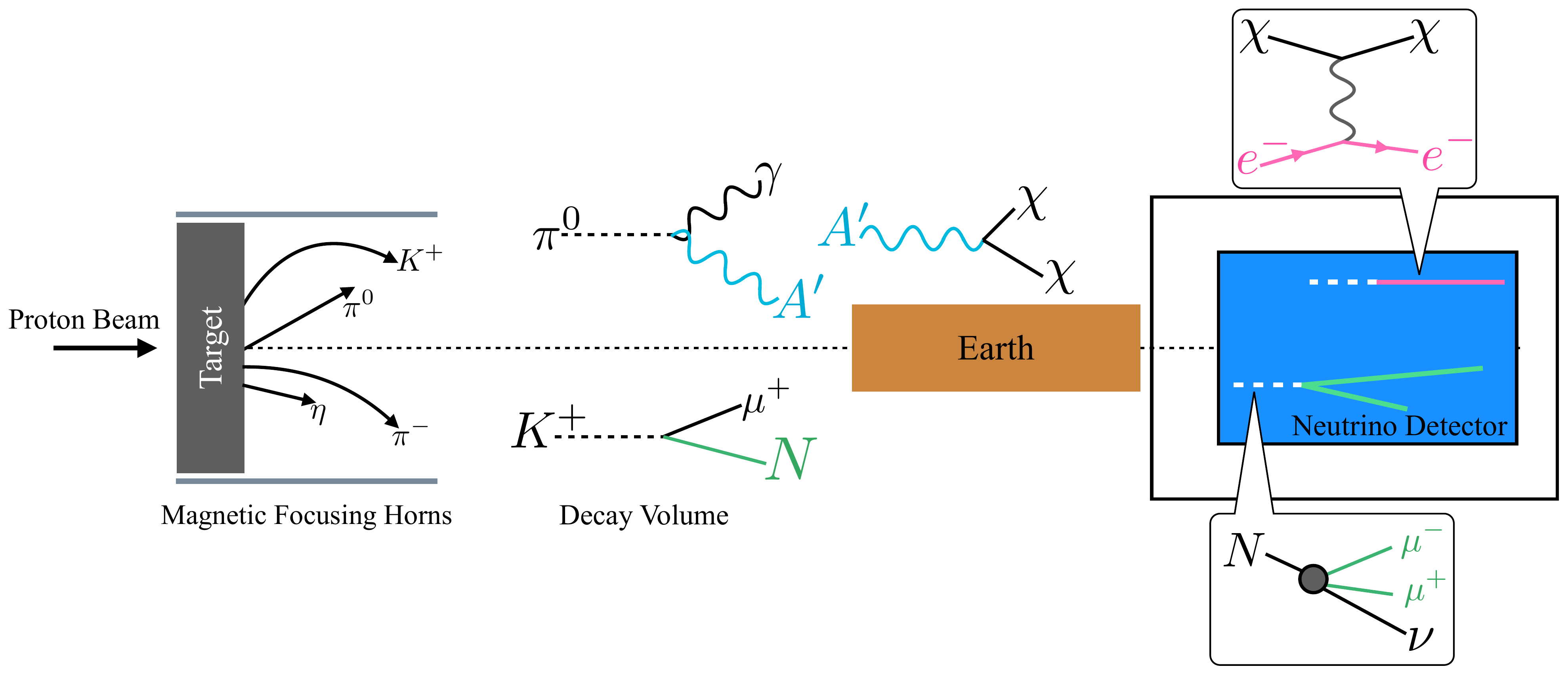}
	\caption{Schematic depicting two classes of beyond-the-SM searches possible in neutrino facilities, specifically those sourced by high-intensity proton beams. The top option demonstrates the production of light dark matter via the decay chain $\pi^0 \to \gamma A^\prime$ (dark photon), $A^\prime \to \chi \bar\chi$ (dark matter), where the dark matter $\chi$ scatters in the detector with an electron and provides a very high-energy, forward-going electron as its signature. The bottom option shows the possible production of a heavy neutral lepton $N$ via charged-kaon decay, where the heavy neutral lepton can travel to, and decay inside the detector via a process such as $N \to \nu \mu^+ \mu^-$.}
	\label{fig:BSMinBeam}
\end{figure}
The general approach, applicable to neutrino facilities sourced by high-energy proton beams, is summarized in~\cref{fig:BSMinBeam}. In these facilities, the focused neutrino beam originates from the production, and subsequent decay of SM particles, specifically charged mesons. For instance, if one desires a pure muon-neutrino beam, the typical approach is to design a set of magnetic focusing horns that selects positively-charged pions in the forward direction so that the eventual $\pi^+ \to \mu^+ \nu_\mu$ results in a large flux of forward-going $\nu_\mu$. At the same time, these magnetic focusing horns \textit{defocus} negatively-charged pions so that the $\pi^- \to \mu^- \bar\nu_\mu$ decay produces muon antineutrinos that are diverted away from the detector(s) in the forward direction. Typically, these focusing horns may have their currents reversed so that the sign selection may be flipped, yielding a pure $\bar\nu_\mu$ beam instead. While these focusing horns are tailored specifically for focusing charged pions, they also have the effect of somewhat focusing/defocusing kaons of the same charge as the desired pions.

These same charged mesons, as well as other SM particles, can instead potentially source the production of new-physics particles. This is shown in the middle portion of~\cref{fig:BSMinBeam}, where a charged meson is shown decaying into a charged muon and a ``heavy neutral lepton'' $N$, a beyond-the-SM particle that is massive and mixes with the SM neutrinos via some small mixing angle, often referred to as $|U_{\mu N}|^2$. In this scenario, as long as $m_N < m_{K^\pm} - m_{\mu}$, the $N$ may be produced in such decays and, up to factors pertaining to the HNL mass relative to the kaon/muon masses, will result in a branching ratio $\mathrm{Br}(K^\pm \to \mu^\pm N) \propto |U_{\mu N}|^2$. Benefitting from the focusing of the charged mesons, these heavy neutral leptons are themselves focused in the forward direction as well. In general, heavy neutral leptons are unstable particles, with their partial widths into SM final states proportional to the same $|U_{\mu N}|^2$. This means that these heavy neutral leptons, on trajectories that intersect with the same neutrino detectors, can decay inside the detector and be identified. We refer the interested reader to, for example, Refs.~\cite{T2K:2019jwa,Kelly:2021xbv,Arguelles:2021dqn,MicroBooNE:2022ctm,MicroBooNE:2023eef} for further discussion of searches for heavy neutral leptons in these facilities, specifically the T2K (Tokai to Kamioka) and MicroBooNE experiments.

A second beyond-the-SM scenario accessible in these environments is also depicted, schematically, in~\cref{fig:BSMinBeam}. This is the concept of a light dark matter (LDM) search, in which dark-matter particles $\chi$ couple to the SM via a dark photon $A^\prime$, which has substantial enough interaction with the SM to be produced in these beam environments. One such scenario is where the dark photon mixes kinetically with the SM photon and thereby can be produced in two-body meson decays, e.g.\ $\pi^0 \to \gamma A^\prime$. If the $A^\prime$ is more than twice as massive as the $\chi$, it can subsequently decay on-shell to $\chi\bar\chi$ pairs (in this description, we are implicitly assuming that $\chi$ is a fermion charged under some dark $U(1)$ gauge charge -- the DM particle can in principle be a scalar particle without loss of generality). Those DM particles will be highly boosted (albeit not as focused as particles emerging from two-body charged-meson decays such as neutrinos and heavy neutral leptons), and a substantial flux will pass through the neutrino detectors. Via the same $A^\prime$ by which it was produced, the $\chi$ can interact with SM particles in the target, for instance nuclei or electrons. DM-electron elastic scattering, kinematically very similar to neutrino-electron scattering, would yield high-energy, very forward-going single-electron signatures. These processes may be searched for in addition to the relatively low SM neutrino-electron elastic scattering event rates in a wide variety of detectors; null results in searches of this type considering the LSND~\cite{deNiverville:2011it,Jordan:2018gcd} and MiniBooNE~\cite{MiniBooNEDM:2018cxm} experimental facilities dominate the exclusion of this model's parameter space for a large range of DM masses. More exotic searches, including ``dark trident'' events with multiple outgoing charged leptons~\cite{deGouvea:2018cfv}, are currently underway in liquid argon time-projection chambers~\cite{MicroBooNE:2023gmv}.

\section{Conclusions}
\label{sec:conclusions}
As neutrino facilities enter the precision era, better and better measurements offer the possibility of discovering new physics. As discussed in this chapter, this includes both new physics associated with neutrinos (e.g.\ through modifications of neutrino oscillations beyond the standard three-flavor paradigm or via new neutrinophilic mediators) or otherwise (including searches for new dark sectors at neutrino facilities). Searches for these types of new physics are underway (and some classes, especially searches for sterile neutrinos, are themselves very mature) and will grow more and more sophisticated in the coming decades.

While looking forward to the future, it is prudent to ask what comes next. It is expected that, over time, researchers will continue developing strategies that maximize the discovery potential of these and other experimental facilities, improving the chances of novel discoveries. If such a discovery is made, then we as a community must capitalize and determine what comes next, including categorizing and classifying the new phenomena. Surely, with such rich experimental facilities on the horizon, this will be an exciting adventure.

\section*{Acknowledgements}
I thank Baha Balantekin for the invitation to participate in this review and for the valuable feedback provided on this draft. The work of KJK is supported in part by US DOE Award \#DE-SC0010813.

\bibliographystyle{Numbered-Style} 
\bibliography{reference}

\end{document}